\documentclass[tightenlines,preprint,showpacs,showkeys,preprintnumbers,amsmath,amssymb]{revtex4}

\usepackage{graphicx}

\newcommand{\Htil}{\tilde{H}}
\newcommand{\Itil}{\tilde{I}}
\newcommand{\p}{\partial}
\newcommand{\half}{\frac{1}{2}}

\newcommand{\thalf}{\tfrac{1}{2}}
\newcommand{\tfourth}{\tfrac{1}{4}}
\newcommand{\al}{\alpha}
\newcommand{\de}{\delta}
\newcommand{\ep}{\epsilon}
\newcommand{\ga}{\gamma}

\newcommand{\om}{\omega}
\newcommand{\si}{\sigma}
\newcommand{\De}{\Delta}
\newcommand{\Ga}{\Gamma}

\newcommand{\vA}{\vec{A}}
\newcommand{\vB}{\vec{B}}

\newcommand{\vk}{\vec{k}}
\newcommand{\vp}{\vec{p}}
\newcommand{\vv}{\vec{v}}
\newcommand{\vx}{\vec{x}}

\newcommand{\etal}{\textit{et al.} }
\newcommand{\yhat}{\hat{y}}
\newcommand{\zhat}{\hat{z}}
\newcommand{\rarr}{\rightarrow}
\newcommand{\rp}{\right)}
\newcommand{\lp}{\left(}
\newcommand{\rb}{\right]}
\newcommand{\lb}{\left[}

\newcommand{\Hbar}{{\bar{H}}}
\newcommand{\Ibar}{{\bar{I}}}
\newcommand{\pbar}{{\bar{p}}}
\newcommand{\qbar}{{\bar{q}}}
\newcommand{\xbar}{{\bar{x}}}
\newcommand{\vbar}{{\bar{v}}}
\newcommand{\zbar}{{\bar{z}}}
\newcommand{\phibar}{{\bar{\phi}}}
\newcommand{\rhobar}{{\bar{\rho}}}
\newcommand{\psibar}{{\bar{\psi}}}
\newcommand{\sech}{\textrm{sech }}

\begin{document}
\title{Coherent acceleration of magnetized ions by electrostatic waves with arbitrary wavenumbers}
\author{D. J. Strozzi}
\email{dstrozzi@mit.edu}
\author{A. K. Ram}
\author{A. Bers}
\affiliation{Plasma Science and Fusion Center, Massachusetts Institute of Technology, Cambridge, Massachusetts 02139}

\begin{abstract}

This paper studies the coherent acceleration of ions interacting with two electrostatic waves in a uniform magnetic field $\vB_0$.  It generalizes an earlier analysis of waves propagating perpendicularly to $\vB_0$ to include the effect of wavenumbers along $\vB_0$.  The Lie transformation technique is used to develop a perturbation theory describing the ion motion, and results are compared with numerical solutions of the complete equations of motion.  Coherent energization occurs when the Doppler-shifted wave frequencies differ by nearly an integer multiple of the ion cyclotron frequency.  When the difference in the parallel wavenumbers of the two waves is increased the coherent energization of ions is limited to a small part of the phase space.  The energization of ions and its dependence on wave parameters is discussed.

\end{abstract}

\pacs{05.45.-a, 45.50-j, 52.20.Dq, 52.50.Sw}
\keywords{wave-particle interactions, coherent ion energization, Lie-transform method}

\date{\today}

\maketitle


\section{Introduction}
The motion of charged particles in the presence of electromagnetic waves is a rich dynamical system that has been studied for a variety of cases.  Important physical applications of this problem occur in laboratory and space plasmas, such as for high-temperature (collisionless) plasma heating and current drive and the transverse energization of ions for times short compared to collisional times.  A particular area of interest is the nonlinear heating of ions (as opposed to linear mechanisms such as Landau and cyclotron damping) by electrostatic waves propagating through a plasma in a uniform magnetic field $\vB_0$.

For a single electrostatic wave propagating across $\vB_0$, the stochastic heating of ions by waves with frequency $\om\gg\om_{ci}$ but $\om\neq N\om_{ci}$ (where $N$ is an integer and $\om_{ci}\equiv qB_0/M$ is the ion cyclotron frequency) was studied by Karney and Bers \cite{karnbers,karney1}.  It was  found that ions with speeds across $\vB_0$ less than the phase velocity of the wave $\om/k_\perp$ (that is, $k_\perp\rho_i\gtrsim\om/\om_{ci}$) exhibit regular motion and do not gain energy.  However, for wave amplitudes above a threshold amplitude, the ions are stochastically heated if their speeds are inside a region with a lower bound near $\om/k_\perp$.    The stochastic ``webs'' generated by a single perpendicularly propagating wave with frequency $\om=N\om_{ci}$ also lead to stochastic ion heating \cite{zaslavsky_weakchaos,fukuyama,chia1}.

For a single wave propagating obliquely to $\vB_0$ it was found that ions could also be stochastically heated \cite{smithpof,smithprl}.  It has recently been shown that single and multiple drift-Alfv\'en waves with $\om<\om_{ci}$ can induce stochastic ion heating \cite{chenliu}, which may account for certain experimental observations \cite{mcchesney}.  For two waves propagating obliquely to $\vB_0$, the threshold wave amplitudes needed for stochastic motion can be significantly lowered \cite{benkadda}.  There is still a lower bound for the stochastic region of phase space.

For two perpendicular waves that satisfy the resonance condition $\om_1-\om_2=N\om_{ci}$ Ram \etal discovered numerically \cite{ram_agu_97} that coherent (as opposed to chaotic) energization can bring ions from low energies into the stochastic domain. B\'{e}nisti \etal \cite{benisti1} then showed that this coherent energization was described by perturbation analysis using Lie transformation methods. The coherent energization was also shown by Ram \etal \cite{abhayjgr} to be described by a multiple time scale analysis, and invoked to explain the energization of hydrogen and oxygen ions from Earth's upper auroral ionosphere into the magnetosphere.  For two non-collinear, perpendicularly propagating waves, the coherent energization was found to
persist as long as the angle between them was less than $30^\circ$ \cite{ram_agu_98}.

Coherent acceleration by electrostatic waves with $\om_1-\om_2=N\om_{ci}$ can only occur when both wave frequencies are larger than $\om_{ci}$.  This process is most interesting for cases where ions with energy well below the stochastic region ($k_\perp\rho_i\ll\om/\om_{ci}$) are accelerated into it.  Most of the work on coherent acceleration has focused on waves with frequencies much higher than $\om_{ci}$.  In magnetic fusion experiments and in the Earth's ionosphere, lower-hybrid waves fit this description ($\om_{lh}\sim\om_{pi}\gg\om_{ci}$, $\om_{lh}=$ lower-hybrid frequency, $\om_{pi}=$ ion plasma frequency).

In this paper we study the interaction of ions with electrostatic waves ranging in frequency from lower-hybrid frequencies down to a few multiples of $\om_{ci}$.  The analysis of B\'enisti \etal \cite{benisti1}  is generalized to include nonzero wavenumbers along $\vB_0$.  We develop a perturbation theory using the Lie transformation method and find conditions for which coherent acceleration persists.  We also discuss the dependence of the range of energization and period of coherent oscillations on wave parameters.

The Hamiltonian formulation of the problem is given in Section II.  An analytic perturbation theory for the coherent motion based on the Lie transformation technique is described in Section III.  Section IV compares the results for the perturbation theory with numerical results obtained from the complete dynamical equations. The scalings of coherent energization and the period of oscillation, for perpendicularly propagating waves, are obtained.  Section V discusses the case of obliquely propagating waves and compares the results with those for two perpendicularly propagating waves.

\section{Equations of Motion}

The nonrelativistic equation of motion of an ion in the presence of a uniform magnetic field $\vB_0=B_0\zhat$ in a plasma and interacting with two electrostatic waves is
\begin{equation}
M\frac{d^2\vx}{dt^2}=q\sum_{i=1}^2 \Phi_i\vk_i\sin(\vk_i\cdot\vx-\om_it+\al_i) + q\vv\times\vB_0
\end{equation}
where $\Phi_i$ is the electrostatic potential amplitude, $\vk_i$ is the wavevector, $\om_i$ is the wave frequency, and $\al_i$ is the phase of the $i^\textrm{th}$ wave.  We normalize times to the inverse of the ion cyclotron frequency $\om_{ci}$, distances to the inverse of $k_{1x}$, and masses to the ion mass $M$.  We restrict our attention to the case where both $\vk_i$'s lie in the $x-z$ plane.  Let $\nu_i\equiv\om_i/\om_{ci}$ and $\ep_i\equiv(\om_{Bi}/\om_{ci})^2$, where $\om_{Bi}\equiv (qk_{1x}^2\Phi_i/M)^{1/2}$ is the bounce frequency in the $i^\textrm{th}$ wave.  The Hamiltonian for this system is
\begin{equation}
h(\vec{x},\vec{p},t) = \thalf(\vec{p}-\vec{A})^2 + \sum_i\ep_i\cos(\vec{k}_i\cdot\vec{x} - \nu_i t + \al_i)
\end{equation}
where $\vec{A}=B_0x\yhat$ is the vector potential, and $\vp=m\vv+q\vA\rarr \vv+x\yhat$ is the (nondimensional) canonical momentum.

Since $h$ is independent of $y$, we can eliminate the $y$ degree of freedom by making a Galilean transformation to a frame moving in the $\yhat$ direction with speed $p_{y0}=v_{y0}+x_0$ (the subscript 0 refers to a quantity's initial value).  Following Ref.~\cite{Goldstein}, the generating function for the canonical transformation from $(y,p_y)$ to $(y',p_y')$ is $F_2=(y-p_{y0}t)(p_y'+p_{y0})$.  Then $y'=y-p_{y0}t$ and $p_y'=p_y-p_{y0}$, so that $p_{y0}'=0$.  The transformed Hamiltonian (to within a constant) is
\begin{equation}
h(x,z,p_x,p_y',p_z,t)=\thalf\lb p_x^2+p_y'^2+(x-p_{y0})^2+p_z^2\rb + \sum_i\ep_i\cos(\vec{k}_i\cdot\vec{x} - \nu_i t + \al_i)
\end{equation}
Since $\p h'/\p y'=0$, $p_y'$ is independent of time so that $p_y'=0$.  This eliminates the $y'$ degree of freedom from the dynamics.  Replacing $x$ by $x'=x-p_{y0}$ and $p_z$ by $v_z$ gives
\begin{equation}
h'(x',z,p_x,v_z,t)=\thalf(p_x^2+x'^2+v_z^2) + \sum_i\ep_i\cos(k_{ix}x' + k_{iz}z - \nu_it + \al_i)
\end{equation}
where $\al_i+k_{ix}p_{y0}$ is replaced by $\al_i$.

In a frame moving with velocity $u\zhat$ the Hamiltonian remains unchanged except that the wave frequencies are Doppler shifted: $\nu_i\rarr\nu_i-k_{iz}u$.  Without loss of generality we assume that $v_{z0}=0$ and consider the effects of $v_{z0}$ on oblique propagation in Section~\ref{sec:oblique}.

We transform $(x',p_x)$ to action-angle coordinates $(\phi,I)$ using the generating function $F_1=\half x'^2\cot\phi$.  Then $I=\half(v_x^2+x'^2)=\half(v_x^2+v_y^2)$ is the perpendicular kinetic energy, and $\phi=\arctan(x'/v_x)=\arctan(-v_y/v_x)$ is the gyrophase.  The transformed Hamiltonian is
\begin{equation}
\label{eqn:fullham}
H(\phi,z,I,v_z,t) = I+\thalf v_z^2+\sum_i\ep_i\cos(k_{ix}\rho\sin\phi+k_{iz}z-\nu_it+\al_i)
\end{equation}
where $\rho=\sqrt{2I}$ is the ion gyroradius.

\section{Perturbation Analysis of Coherent Motion}
\label{sec:pert}
\begin{figure}[p]
\includegraphics{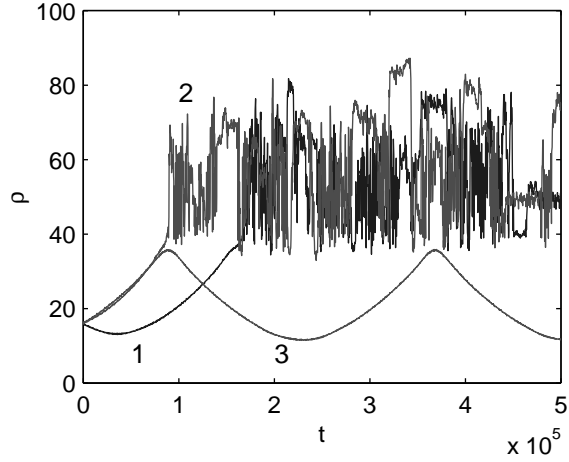}
\caption{\label{fig:rvst_perp_full} $\rho$ versus $t$ for three ions interacting with two perpendicular waves from the full Hamiltonian $H$ (\ref{eqn:fullham}).  Quantities in all figures are given in terms of the normalized units defined in the text.  The initial $\rho_0=15.95\ (\xi_0=\rho_0/\nu_1=0.4)$ for all three ions while their phases are $\phi_0=(-0.3,0.2,0.4)\pi$ for ions labelled 1, 2, and 3, respectively.  The parameters for the two waves are: \mbox{$\ep_1=\ep_2=4$,} \mbox{$k_{1x}=k_{2x}=1$,} \mbox{$k_{1z}=k_{2z}=0$,} $\nu_1=40.37$, and $\nu_2=\nu_1-1$.}
\end{figure}

\begin{figure}[p]
\includegraphics{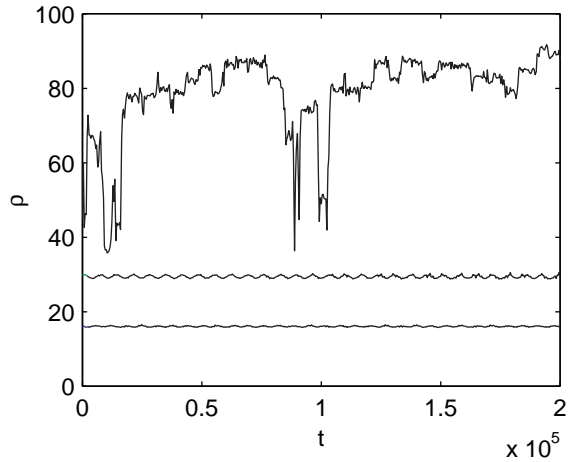}
\caption{\label{fig:rvst_perp_off} $\rho$ versus $t$ for the same parameters as in Fig.~\ref{fig:rvst_perp_full} except that $\rho_0=15.95,30,45$ and \mbox{$\nu_2=\nu_1-1.001$.}}
\end{figure}

In general, the equations of motion obtained from (\ref{eqn:fullham}) cannot be solved analytically.  Consequently, we resort to numerical solutions to provide an insight into the dynamics of ions in two electrostatic waves.  Figure~\ref{fig:rvst_perp_full} shows the time evolution of $\rho$ for three ions having the same initial $\rho_0$, but different $\phi_0$, interacting with two waves of frequencies $\nu_1=40.37$ and $\nu_2=39.37$, and amplitudes $\ep_1=\ep_2=4$.  (All the numerical solutions of ordinary differential equations have been carried out using the Bulirsch-Stoer algorithm described in Ref.~\cite{numrec}.)  There are two distinct kinds of motion: the slow, smooth, ``coherent'' oscillations at lower $\rho$, and the irregular, ``stochastic'' motion at higher $\rho$.  Superimposed on the coherent motion are small-amplitude, high-frequency fluctuations.  Figure~\ref{fig:rvst_perp_off} shows the orbits for the same parameters as Fig.~\ref{fig:rvst_perp_full} except that $\nu_2=39.369$ and the initial conditions are different.  This demonstrates that the coherent acceleration from low to high energies occurs only when $\nu_1-\nu_2$ is an integer.

Our interest is to provide an analytical description of the coherent dynamics without going into details of the stochastic region, other than to note its existence for $\rho \approx\min(\nu_i)$ \cite{karney1,karnbers}.  We assume that the waves are perturbing the cyclotron motion of the ions and express
\begin{equation}
H=H_0+H_1
\end{equation}
where
\begin{equation} \label{eqn:H0H1}
H_0=I+\thalf v_z^2 \quad\mathrm{and}\quad H_1=\sum_i\ep_i\cos(k_{ix}\rho\sin\phi+k_{iz}z-\nu_it+\al_i).
\end{equation}
An approximate analytical description of the ion motion in the coherent regime is obtained by using the Lie perturbation technique \cite{cary,lichtlieb} with the ordering parameter $\ep\ (\ep\sim\ep_1\sim\ep_2$).  We assume that $\nu_i\notin\mathcal{Z}$ but $(\nu_1-\nu_2)=N\in\mathcal{Z}$.  For $\nu_i\in\mathcal{Z}$ a web structure is formed in phase space and has been discussed elsewhere for a single wave \cite{fukuyama} and for two waves propagating across $\vB_0$ \cite{benisti2}.  For the case of a single wave the stochastic web structure also has a lower bound \cite{benisti_pla}.

From the Lie perturbation analysis (Appendix \ref{app:lie}) the Hamiltonian that describes the coherent ion motion to $O(\ep^2)$ is
\begin{equation}
\Hbar(\phibar,\zbar,\Ibar,\vbar_z,t) = \Ibar+\thalf \vbar_z^2+\Hbar_2
\end{equation}
where
\begin{eqnarray}
\Hbar_2  &=& S_0(\Ibar,\vbar_z)+S_-(\Ibar,\vbar_z)\cos(N(\phibar-t)+\De k_z\zbar+\al_1-\al_2)
\\S_0    &=& S_{0x} + S_{0z}    \label{eqn:S0}
\\S_{0x} &=&-\frac{1}{2\rhobar}\sum_i k_{ix}\ep_i^2\frac{m}{m-\mu_i}J_{m,i}J_{m,i}'
\\S_{0z} &=& \tfourth \sum_i k_{iz}^2\ep_i^2 \frac{J_{m,i}^2}{(m-\mu_i)^2}  \label{eqn:S0z}
\\S_-    &=& S_{-x} + S_{-z}    \label{eqn:Sm}
\\S_{-x} &=& -\frac{\ep_1\ep_2}{4\rhobar(m-\mu_1)}(k_{1x}(m-N)J_{m,1}'J_{m-N,2}+k_{2x}mJ_{m,1}J_{m-N,2}')
\\ \nonumber &&  -\frac{\ep_1\ep_2}{4\rhobar(m-\mu_2)}(k_{1x}mJ_{m,2}J_{m+N,2}'+k_{2x}(m+N)J_{m,2}'J_{m+N,1})
\\S_{-z} &=& \tfourth k_{1z}k_{2z}\ep_1\ep_2 \lp \frac{J_{m,1}J_{m-N,2}}{(m-\mu_1)^2} + \frac{J_{m,2}J_{m+N,1}}{(m-\mu_2)^2} \rp \label{eqn:Smz}
\end{eqnarray}
$\De k_z=(k_{1z}-k_{2z}), \mu_i=\nu_i-k_{iz}\vbar_z,$ and $m$ is summed from $-\infty$ to $+\infty$.  $J_{m,i}\equiv J_m(k_{ix}\rhobar)$ is the Bessel function of the first kind and $f'(\xi)=df/d\xi$.  The barred coordinates are related to the original coordinates by a near-identity transformation: $(\phibar,\zbar,\Ibar,\vbar_z)=(\phi,z,I,v_z)+O(\ep)$ (Appendix \ref{app:lie}).  For instance, the relation between $I$ and $\Ibar$ is:
\begin{equation}
\label{eqn:ItoIbar}
I \approx \Ibar-\ep_i\sum_m\frac{mJ_{m,i}}{m-\mu_i}\cos(m\phibar+k_{iz}\zbar-\nu_it+\al_i)
\end{equation}
The Hamiltonian $\Hbar$ is a generalization to oblique waves of the results obtained in \cite{benisti1} for collinear perpendicularly propagating waves.  In the limit $k_{iz}\rarr0$ the above reduces to the description in Ref.~\cite{benisti1}.  A nonzero $\al_1-\al_2$ is equivalent to a shift in the initial $\phibar_0$ so that, without loss of generality, we can set $\al_1=\al_2=0$.

Our perturbation analysis assumes there are no resonances at $O(\ep)$.  Such resonances occur if $\nu_i$ is an integer, where our present analysis breaks down.

The explicit time dependence in $\Hbar$ can be eliminated by transforming from $\phibar$ to $\psibar=\phibar-t$ using the generating function $F_2=\tilde{I}(\phibar-t)$.  The transformed Hamiltonian is:
\begin{equation}
\label{eqn:Htil}
\Htil(\psibar,\zbar,\Ibar,\vbar_z) = \thalf \vbar_z^2 + S_0(\Ibar,\vbar_z) + S_-(\Ibar,\vbar_z)\cos(N\psibar+\De k_z\zbar)
\end{equation}
where $\Ibar=\Itil$ has replaced $\Itil$ ($\Ibar$ and $\psibar$ are canonically conjugate).  Since $\Htil$ does not depend explicitly on time it is a constant of the motion.

Using Hamilton's equations for $\dot{\Ibar}$ and $\dot{\vbar}_z$, we find a second constant of the motion:
\begin{equation} \label{eqn:vzconst} \frac{d}{dt}\lp \vbar_z-\frac{\De k_z}{N}\Ibar\rp=0 \end{equation}
Thus, the system is integrable and the dynamics described by $\Htil$ are not stochastic.  Along an orbit, $\vbar_z$ is a function of $\Ibar$ and initial conditions only:
\begin{equation} \label{eqn:vzlie}
\vbar_z = v_{z0}+\frac{\De k_z}{N}(\Ibar-I_0).
\end{equation}
Therefore, $S_0$ and $S_-$ are functions just of $\Ibar$.  Since $|\cos{x}|\leq1$
\begin{equation}
\label{eqn:Hbound}
H_-\leq \tilde{H} \leq H_+, \qquad H_\pm(\Ibar)=\thalf \vbar_z^2 + S_0(\Ibar)\pm |S_-(\Ibar)|
\end{equation}
For an initial condition with a given value of $\Htil$, $\rhobar$ varies between the two points where $\Htil$ equals $H_+(\rhobar)$ or $H_-(\rhobar)$.  We refer to $H_\pm$ as the potential barriers, since they delimit the allowed and forbidden regions of phase space.

Figure~\ref{fig:rvst_perp_coh} shows the orbits generated by the second-order Hamiltonian (\ref{eqn:Htil}) for the same parameters  as in Fig.~\ref{fig:rvst_perp_full}.  Our perturbation analysis accurately captures the coherent motion of the full system except near the stochastic region $\rho\approx\min(\nu_i)$ where our perturbation theory breaks down.  Below this region, $\rho$ and $\rhobar$ differ by small fluctuations that are accounted for, to $O(\ep)$, by the transformation (\ref{eqn:ItoIbar}).

\section{Coherent Motion for Perpendicular Waves}
\begin{figure}[p]
\includegraphics{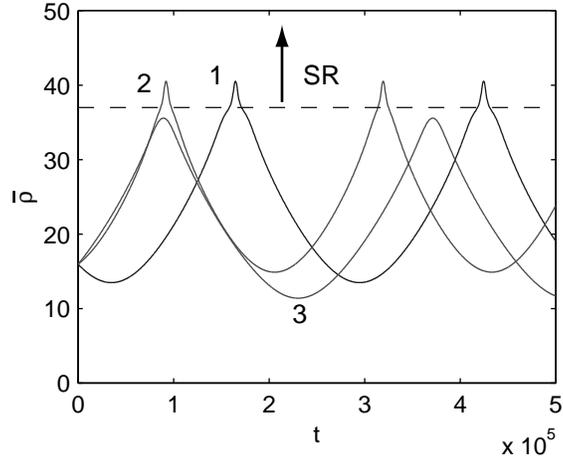}
\caption{\label{fig:rvst_perp_coh} $\rhobar$ versus $t$ from the coherent Hamiltonian $\Htil$ (\ref{eqn:Htil}) for the same parameters as in Fig.~\ref{fig:rvst_perp_full}.  SR indicates the stochastic region for the full Hamiltonian.}
\end{figure}

\begin{figure}[p]
\includegraphics{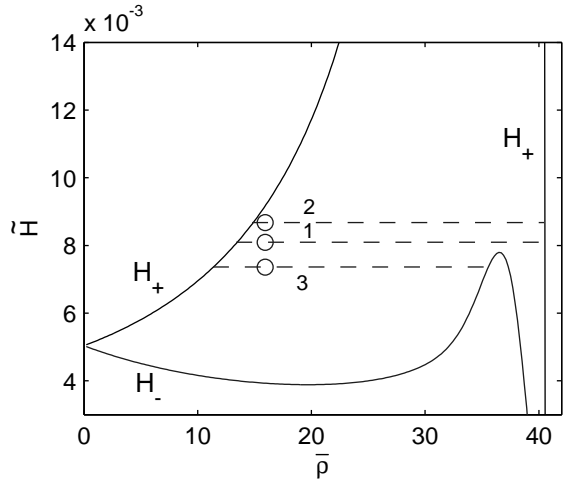}
\caption{\label{fig:hvsr_perp} $H_+$ and $H_-$ versus $\rhobar$ for the same parameters as in Fig.~\ref{fig:rvst_perp_full}.  The initial values of $\Htil$ for the three ions in Fig.~\ref{fig:rvst_perp_full} are marked by the open circles.}
\end{figure}

Using the Hamiltonian (\ref{eqn:Htil}) we now analyze the ion motion for two perpendicularly propagating waves. Figures~\ref{fig:rvst_perp_full} and \ref{fig:rvst_perp_coh} show the complete and coherent motion, respectively, for two perpendicularly propagating waves ($k_{iz}=0$).  Figure~ \ref{fig:hvsr_perp} displays $H_+$ and $H_-$ from (\ref{eqn:Hbound}) for the same parameters as in Figs.~\ref{fig:rvst_perp_full} and \ref{fig:rvst_perp_coh}, and the values of $\Htil$ for the three initial conditions.  The coherent analysis correctly predicts that particle 3 in Fig.~\ref{fig:rvst_perp_full} will not make it into the chaotic regime because it is reflected by the bump in $H_-$.

If we multiply $\ep_1$ and $\ep_2$ by the same factor $a$ then $\Htil$ in (\ref{eqn:Htil}) is multiplied by $a^2$ (note that for perpendicularly propagating waves, $\thalf \vbar_z^2$ is a constant and can be eliminated from $ \Htil$).  Since a rescaling of the Hamiltonian is equivalent to a rescaling of time, rescaling both $\ep_i$'s does not affect the range of motion in $\rhobar$ but rescales the period by $1/a^2$.  For $\ep_1\sim\ep_2$, this means the period scales like $1/\ep_1^2$.  This reflects the fact that the coherent motion is second-order in the wave field amplitudes.  It also shows that in certain physical situations, at sufficiently large amplitudes, the effects of collisions on the coherent energization can be made negligible.

\begin{figure}[p]
\includegraphics{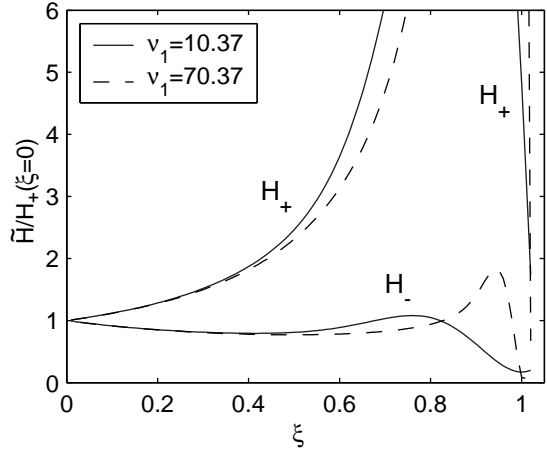}
\caption{\label{fig:hvsr_nu} $H_\pm/H_+(\xi=0)$ versus $\xi\equiv\rhobar/\nu_1$ for $N=1$ and $\nu_1=10.37$ (solid line) and $\nu_1=70.37$ (dashed line).  $\ep_1=\ep_2=$arbitrary, $k_{1x}=k_{2x}=1$, and $k_{1z}=k_{2z}=0$.}
\end{figure}

\begin{figure}[p]
\includegraphics{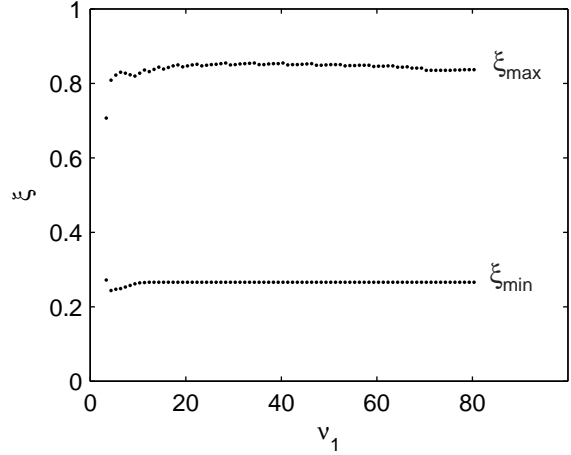}
\caption{\label{fig:xirng_nu} Average $\xi_{min}$ and $\xi_{max}$ versus $\nu_1$ for perpendicularly propagating waves based on the barriers $H_\pm$.  Parameters are as in Fig.~\ref{fig:hvsr_nu} except that $\nu_1=(3.37,4.37,...,80.37), \xi_0=0.4$, and the average is over $\phi_0=(0,0.05,...,1)\pi$.}
\end{figure}

The range of coherent motion in $\rho$ scales linearly with the wave frequencies.  In Fig.~\ref{fig:hvsr_nu} we plot $H_\pm/H_+(\xi=0)$ versus $\xi\equiv\rhobar/\nu_1$ for two values of $\nu_1$ with $N=1$.  Note that the potential barriers do not change significantly with $\nu_1$.  Figure~\ref{fig:xirng_nu} shows that, as a function of $\nu_1$, the average $\xi_{min}$ and $\xi_{max}$ have a small variation ($\xi_{min}$ and $\xi_{max}$ are the maximum and minimum $\xi$ attained by an ion undergoing coherent motion and occur when the ion reaches the barriers $H_\pm$).  The average is over ions with the same initial $\xi_0=0.4$ and different \mbox{$\phi_0=(0,0.05,\ldots,1)\pi$.}  Waves with higher frequencies can therefore produce coherent energization to higher energies.  Since the lower bound of the stochastic region is \mbox{$\rho\approx\min(\nu_i)\rarr\xi\approx1$,} ions with the same initial $\xi_0$ and $\phi_0$ either will or will not reach the stochastic region regardless of the wave frequencies.  For $\nu_1$ near an integer $\xi_{max}$ is about 20\% higher than when $\nu_1$ is near a half-integer (this is not shown explicitly in Fig.~\ref{fig:xirng_nu}).

\begin{figure}[p]
\includegraphics{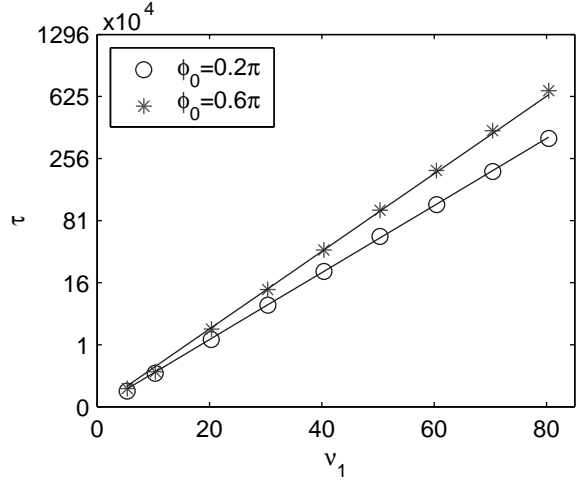}
\caption{\label{fig:period} The period of coherent oscillation $\tau$ versus $\nu_1$.  The other wave parameters are the same as in Fig.~\ref{fig:rvst_perp_full}.  The initial conditions are $\xi_0=0.4$ and $\phi_0=(0.2,0.6)\pi$. The open circles and stars are the periods obtained from integrations of the dynamics given by $\Htil$.  The solid lines are proportional to $\nu_1^4$ with the constant of proportionality chosen to match the period at $\nu_1=40.37$.  The vertical axis is scaled so that $\nu_1^4$ is a straight line.}
\end{figure}

The period of oscillation in $\Ibar$ (see Fig.~\ref{fig:rvst_perp_coh}) can be estimated from the equation of motion for $\dot{\Ibar}$:
\begin{equation}
\label{eqn:idot} \dot{\Ibar}=-\frac{\p\Htil}{\p\psibar}=NS_-\sin(N\psibar)
\end{equation}
An orbit's turning point typically occurs when $\psibar=n\pi/N$, i.e., when it hits one of the barriers $H_\pm$.  Therefore, approximately, the period of oscillation $\tau$ is given by $\tau \approx 2\pi/(N\langle\dot{\psibar}\rangle)$, where $\langle\rangle$ denotes the average over one period.  From the asymptotic forms of $S_0$ and $S_-$ for $\nu_1\sim\nu_2\gg1$ (Appendix \ref{app:asym}) we find that
\begin{equation}
\Htil\approx \nu_1^{-2}h_a(\xi,\psibar)
\end{equation}
where $h_a$ depends on $\nu_i$ only through $\xi$.  Then
\begin{equation}
\label{eqn:psidot} \dot{\psibar} = \frac{\p\Htil}{\p\Ibar} \approx \nu_1^{-2}\frac{\p h_a}{\p\Ibar}=\nu_1^{-4}\frac{1}{\xi}\frac{\p h_a}{\p\xi}
\end{equation}
Thus, $\tau \sim \nu_1^4$.  Waves of lower frequency accelerate ions much more rapidly than those with higher frequency and thus may also be made less sensitive to the effects of collisions.  Figure~\ref{fig:period} compares this scaling with the periods of two actual orbits obtained from $\Htil$.

\section{Coherent Motion for Oblique Waves}
\label{sec:oblique}
In this section we describe the motion of ions when the waves have nonzero parallel wavenumber $k_{iz}$.  This couples the parallel dynamics to the perpendicular motion.

For ions with initial $v_{z0}=0$ interacting with a \textit{single} oblique wave, the motion is stochastic when \cite{smithpof}:
\begin{equation}
\sqrt{|J_{n_0}(\rho)|}+\sqrt{|J_{n_0+1}(\rho)|} \geq \frac{1}{2k_z\sqrt{\ep}}
\end{equation}
where $n_0$ is the greatest integer less than $\nu$.  For $n_0\gg1$, the lower bound of the stochastic region is
\begin{equation}
\rho\approx n_0+\frac{0.15}{\ep k_z^2}n_0^{2/3}-1.1n_0^{1/3}
\end{equation}
As for a single perpendicular wave, the lower bound in $\rho$ is roughly the wave frequency and decreases with $\ep$.  The stochastic region in $v_z$ extends from $v_z\approx0$ to $v_z\approx2\nu$.

For \textit{two} waves Eq. (\ref{eqn:vzlie}) shows that $\vbar_z$ changes only when $\De k_z\neq0$.  Thus, the cases $\De k_z=0$ and $\De k_z\neq0$ lead to different dynamics and are treated separately.

\subsection{Equal Parallel Wavenumbers: $\De k_z=0$}
\begin{figure}[p]
\includegraphics{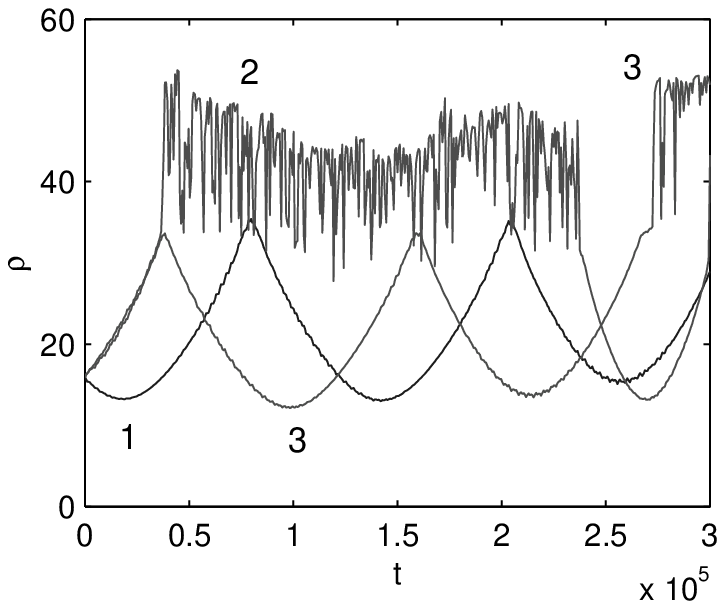}
\caption{\label{fig:rvst_dkz0} $\rho$ versus $t$ for the same parameters as in Fig.~\ref{fig:rvst_perp_full} except that $k_{1z}=k_{2z}=1$.}
\end{figure}

\begin{figure}[p]
\includegraphics{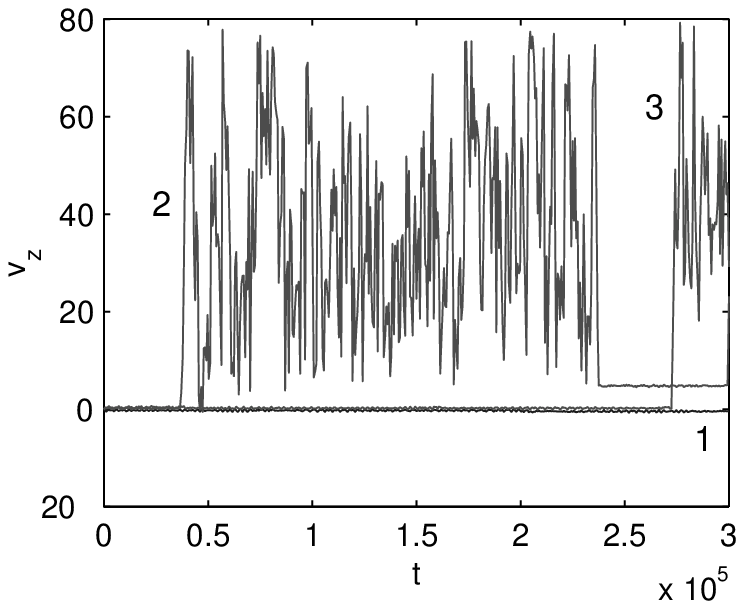}
\caption{\label{fig:vzvst_dkz0} $v_z$ versus $t$ for the same parameters as in Fig.~\ref{fig:rvst_dkz0}.}
\end{figure}

Figures~\ref{fig:rvst_dkz0} and \ref{fig:vzvst_dkz0} show the time evolution of $\rho$ and $v_z$ for two waves propagating at an angle of $45^\circ$ ($k_{iz}=k_{ix}=1)$ to $\vB_0$.  As in the case of two perpendicularly propagating waves, there is coherent change in $\rho$.  During this coherent evolution $v_z$ has small-amplitude fluctuations around its initial value.  In the region where the motion in $\rho$ becomes stochastic so does the motion in $v_z$.  The stochastic region in $v_z$ agrees with the estimate given above.  Since $\vbar_z$ is a constant during the coherent motion, the fluctuations in $v_z$ are due to the transformation between $\vbar_z$ and $v_z$.

\begin{figure}[p]
\includegraphics{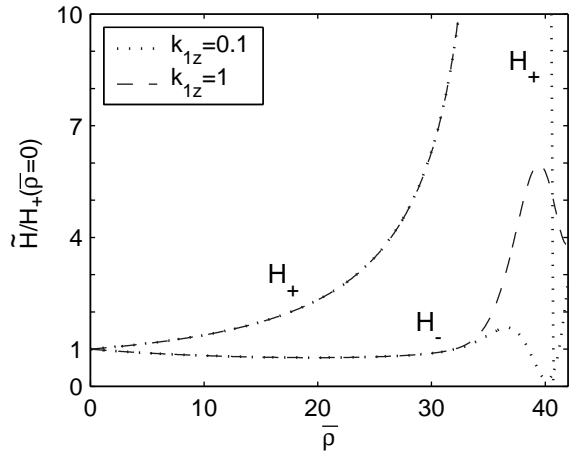}
\caption{\label{fig:hvsr_k1z} $H_\pm/H_+(\rhobar=0)$ versus $\rhobar$ for $k_{1z}=k_{2z}=0.1$ (dots) and 1 (dashes).  Other parameters are as in Fig.~\ref{fig:rvst_dkz0}.  The curves for $k_{1z}=0$ are very close to those for $k_{1z}=0.1$.}
\end{figure}

\begin{figure}[p]
\includegraphics{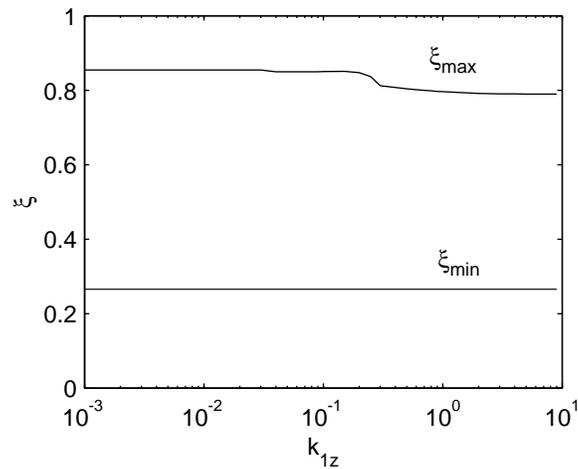}
\caption{\label{fig:xirng_k1z} Average $\xi_{min}$ and $\xi_{max}$ versus $k_{1z}$ for $k_{2z}=k_{1z}$ from $H_\pm$.  $\nu_1=40.37$ and the other parameters are as in Fig.~\ref{fig:xirng_nu}.}
\end{figure}

The main effect of equal parallel wavenumbers is to slightly decrease the range of coherent motion from what it is for perpendicularly propagating waves, thus inhibiting some ions from reaching the stochastic region.  Numerical studies indicate that $|S_{0x}/S_{0z}|$ and $|S_{-x}/S_{-z}|$ are both unity for $\xi<1$ but approach 0 as $\xi\rarr1$.  This raises the bump in $H_-$ as $k_{iz}$ is increased.  Consequently, more ions are reflected by $H_-$ and the range of coherent motion in $\rho$ is slightly lowered.  This is evident from Fig. \ref{fig:hvsr_k1z}, which shows $H_\pm/H_+(\rhobar=0)$ for $k_{1z}=0.1$ and 1.  Figure~\ref{fig:xirng_k1z} shows the range of motion $\xi_{min},\xi_{max}$ for different $k_{1z}$.  Increasing $k_{iz}$ slightly lowers $\xi_{max}$ since the enhanced bump in $H_-$ reflects more ions.  Generally then, significant coherent energization and access to the stochastic region is obtainable with oblique waves provided that $\De k_z = 0$, while the normalized $k_z$ may be large.

\subsection{Unequal Parallel Wavenumbers: $\De k_z\neq0$}
\begin{figure}[p]
\includegraphics{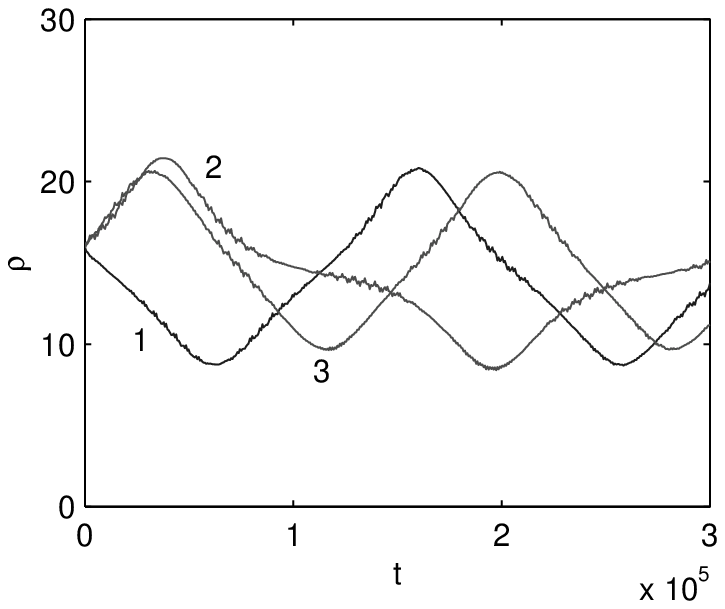}
\caption{\label{fig:rvst_dkzp001} $\rho$ versus $t$ for the same parameters as in Fig.~\ref{fig:rvst_perp_full} except that $k_{1z}=0.001$ and $k_{2z}=0$.}
\end{figure}

\begin{figure}[p]
\includegraphics{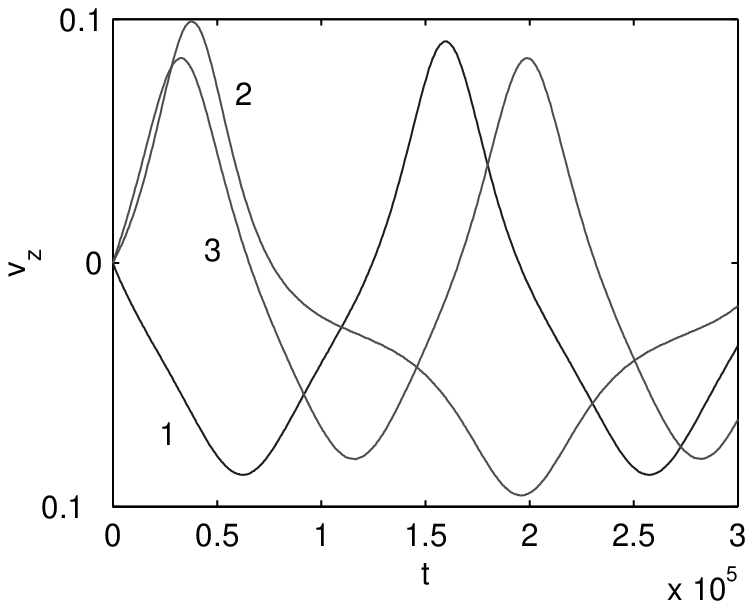}
\caption{\label{fig:vzvst_dkzp001} $v_z$ versus $t$ for the same parameters as in Fig.~\ref{fig:rvst_dkzp001}. }
\end{figure}

When the parallel wavenumbers of the two waves are different, the coherent motion of the ions changes drastically.  In this case $v_z$ undergoes coherent motion and the term $\thalf \vbar_z^2$ in $\Htil$ (\ref{eqn:Htil}) is no longer a constant.  This limits the range in $\rho$ as $\De k_z$ is increased.  Figures~\ref{fig:rvst_dkzp001} and \ref{fig:vzvst_dkzp001} show the time evolution of $\rho$ and $v_z$ for the exact orbits obtained from (\ref{eqn:fullham}) with $k_{1z}=0.001$ and $k_{2z}=0$.  These figures illustrate the limits in $\rho$.

\begin{figure}[p]
\includegraphics{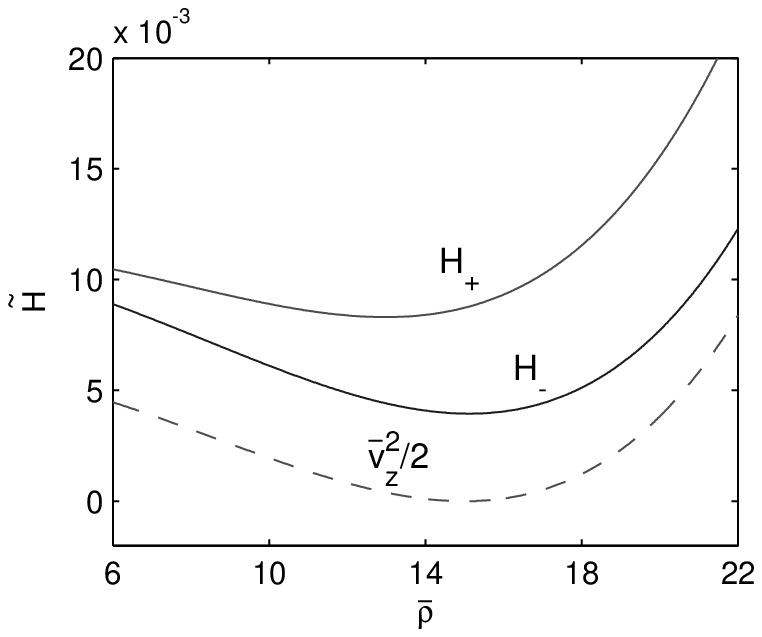}
\caption{\label{fig:hvsr_dkzp001} $H_\pm$ and $\thalf \vbar_z^2$ versus $\rhobar$ for the same parameters as in Fig.~\ref{fig:rvst_dkzp001}. }
\end{figure}

\begin{figure}[p]
\includegraphics{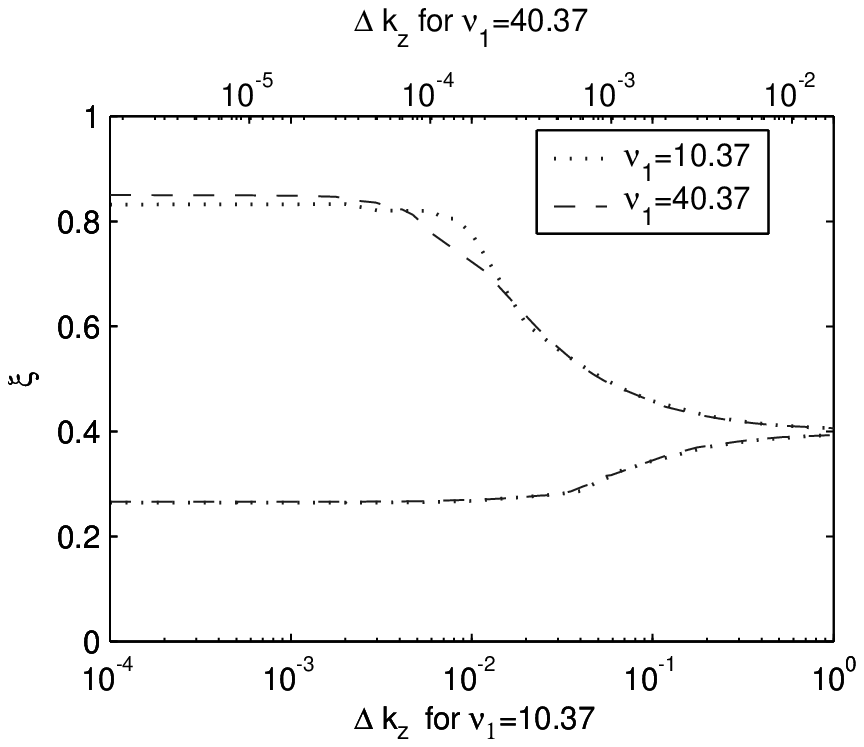}
\caption{\label{fig:xirng_dkz} Average $\xi_{min}$ and $\xi_{max}$ versus $\De k_z$ from $H_\pm$ for $k_{1z}=0.1,\ k_{2z}=k_{1z}-\De k_z,\ \nu_1=10.37$ (dotted) or 40.37 (dashed), and the other wave parameters as in Fig.~\ref{fig:rvst_perp_full}.  Initial conditions and averaging are as in Fig.~\ref{fig:xirng_nu}.  The abscissa for $\nu_1=40.37$ has been rescaled by $(10.37/40.37)^3$.  If the scaling in (\ref{eqn:dkzscal}) were exact, the dotted and dashed lines would coincide.}
\end{figure}

Figure~\ref{fig:hvsr_dkzp001} shows the variation of $H_\pm$ and $\thalf \vbar_z^2$ as functions of $\rhobar$.  For $\rhobar$ far from $\rhobar_0=17$,  $H_+-H_-=2|S_-|\ll\thalf\vbar_z^2$ so that $H_+\approx H_-$.  Figure~\ref{fig:xirng_dkz} shows the limitation on the range of coherent motion in $\xi$ for $\De k_z\neq0$.

The coherent motion in $v_z$ has the effect of detuning the waves from exact resonance.  The resonance condition for an ion with $v_z\neq0$ is
\begin{equation} \label{eqn:rescond}
R \equiv \nu_{10}-\nu_{20}-(\De k_z) v_z\in\mathcal{Z}
\end{equation}
where $\nu_{10}$ and $\nu_{20}$ are the wave frequencies in the laboratory frame, and $\nu_i=\nu_{i0}-k_{iz}v_z$.  $\Htil$ describes the ion's motion as long as $R$ is close to an integer.  For $\De k_z\neq0$, $v_z$ changes coherently.  Condition (\ref{eqn:rescond}) is not satisfied for all times, and the resonant interaction becomes less effective.  The coherent change in $v_z$ thus limits itself, which keeps $R$ close to an integer.  Since the coherent changes in $\rhobar$ and $\vbar_z$ are linked via (\ref{eqn:vzlie}), the coherent change in $\rhobar$ is also small.

Consider a distribution of ions with different initial $v_{z0}$ interacting with two waves of frequencies $\nu_{10}$ and $\nu_{20}$.  For $\De k_z=0$, all ions will be in resonance with the waves provided $\nu_{10}-\nu_{20}\in\mathcal{Z}$.  For $\De k_z\neq0$, the resonance condition (\ref{eqn:rescond}) implies that only ions with certain $v_{z0}$, namely
\begin{equation}
v_{z0} \approx \frac{\nu_{10}-\nu_{20}-n}{\De k_z}, \qquad n\in\mathcal{Z}
\end{equation}
are initially in resonance.  As $v_z$ changes coherently, they fall out of resonance.

\begin{figure}[p]
\includegraphics{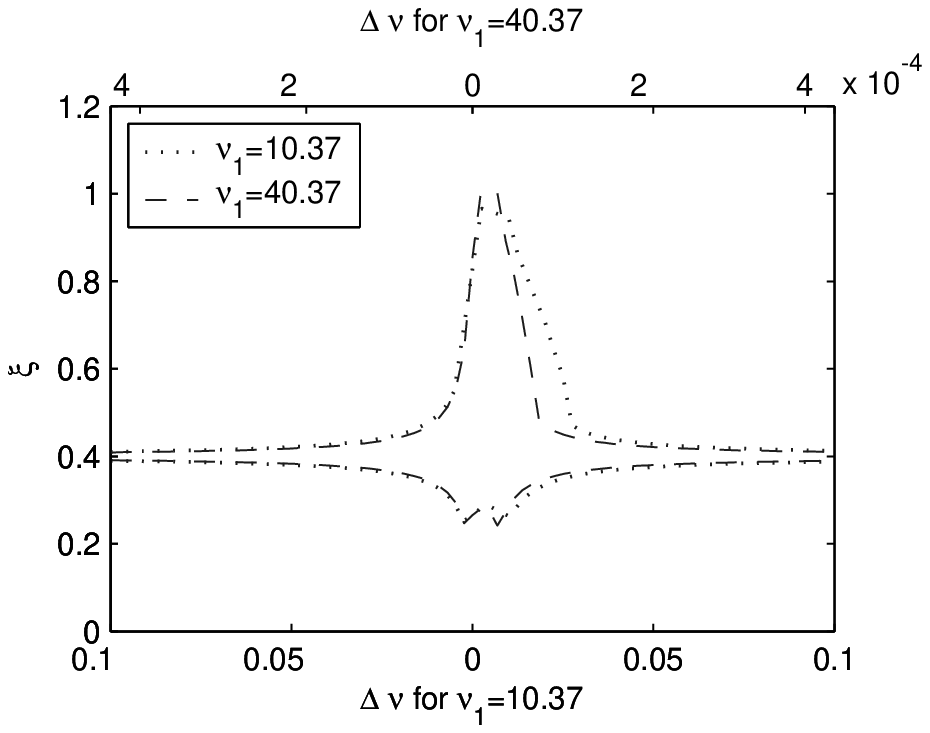}
\caption{\label{fig:xirng_dnu} Average $\xi_{min}$ and $\xi_{max}$ versus $\De\nu$ from $H_\pm$ for $\nu_2=\nu_1-1-\De\nu,\ \nu_1=10.37$ (dotted) or 40.37 (dashed) and the other wave parameters as in Fig.~\ref{fig:rvst_perp_full}.   Initial conditions and averaging are as in Fig.~\ref{fig:xirng_nu}.  The abscissa for $\nu_1=40.37$ has been rescaled by $(10.37/40.37)^4$.  If the scaling in (\ref{eqn:dnuscal}) were exact, the dotted and dashed lines would coincide.}
\end{figure}

This situation is analogous to the case of two perpendicularly propagating waves when the wave frequencies do not differ by an integer \cite{benisti1}.  Following Section IV.C of \cite{benisti1}, the approximate Hamiltonian, correct to second order in wave amplitudes, that describes the coherent motion is
\begin{equation} \label{eqn:hoff}
\Htil_{off}=-\frac{\De\nu}{N}\Ibar + \Htil
\end{equation}
where $(\nu_1-\nu_2)=N+\De\nu$ and $|\De\nu|\ll 1$.  In this case the barriers $H_\pm$ are given by
\begin{equation} \label{eqn:Hpm_off}
H_\pm=-\frac{\De\nu}{N}\Ibar+\thalf \vbar_z^2+S_0\pm|S_-|
\end{equation}
The first term in (\ref{eqn:hoff}) limits the coherent motion, and plays a similar role to $\thalf \vbar_z^2$.  Figure~\ref{fig:xirng_dnu} shows the range of motion in $\xi$ as a function of $\De\nu$.  the largest range of coherent motion occurs for $\De\nu$ slightly different from 0, which allows $-(\De\nu/N)\Ibar$ to partly cancel $S_0$ in \mbox{Eq. (\ref{eqn:Hpm_off}).}

As the wave frequencies are increased, the range in $\De k_z$ and $\De\nu$ for which there is appreciable coherent motion becomes much narrower.  Let $\xi_a(\De k_z)$ be either the upper or lower bound of coherent motion in $\xi$ for wave frequencies $\nu_{1a}$ and $\nu_{2a}=\nu_{1a}-N$.  The asymptotic forms in Appendix~\ref{app:asym} indicate that for two different frequencies $\nu_{1b}$ and \mbox{$\nu_{2b}=\nu_{1b}-N$},
\begin{equation} \label{eqn:dkzscal}
\xi_b(\De k_z) \approx \xi_a\lp \lp\frac{\nu_{1b}}{\nu_{1a}}\rp^3 \De k_z \rp
\end{equation}
Suppose $\xi_a$ is large for $k_1\leq\De k_z\leq k_2$, and that $\nu_{1b}=4\nu_{1a}$.  Then $\xi_b$ is large only for $k_1/64\leq\De k_z\leq k_2/64$. Coherent motion occurs over a smaller range of $\De k_z$ when the wave frequencies are larger.  Similarly,
\begin{equation} \label{eqn:dnuscal}
\xi_b(\De\nu) \approx \xi_a\lp \lp\frac{\nu_{1b}}{\nu_{1a}}\rp^4 \De\nu \rp
\end{equation}
Figures \ref{fig:xirng_dkz} and \ref{fig:xirng_dnu} demonstrate the range of coherent motion versus $\De k_z$ and $\De\nu$, respectively, and validate the scalings in (\ref{eqn:dkzscal}) and (\ref{eqn:dnuscal}) with wave frequency.  As the wave frequencies are increased, $\De k_z$ and $\De\nu$ must be much smaller for ions to be energized to the stochastic region.  Hence, just as in the cases of perpendicular propagation or $\De k_z=0$, for nonzero but small $\De k_z$ energization by waves with low frequencies is more advantageous than by waves with high frequencies.

\section{Conclusions}

We have shown that two electrostatic waves propagating obliquely to an ambient magnetic field can coherently energize ions when their Doppler-shifted frequencies differ by a multiple of the ion cyclotron frequency.  A second-order Hamiltonian, derived using the Lie perturbation technique, accurately describes the coherent motion and agrees well with numerical simulations of the complete dynamical equations.  The energization of ions occurs regardless of the angle of wave propagation provided the parallel wavenumbers of the two waves are approximately equal.  If the parallel wavenumbers are equal, there is no coherent acceleration along $\vB_0$ but considerable stochastic energization both along and across $\vB_0$.  Moreover, the perpendicular coherent motion is quite similar to the case of perpendicularly propagating waves.  There is a small amount of coherent acceleration along $\vB_0$ when the parallel wavenumbers differ, but this causes the resonance condition to be violated.  A difference between the parallel wavenumbers is similar to the difference between $(\om_1-\om_2)/\om_{ci}$ and the nearest integer.

There is no threshold ion energy or wave amplitude required for the coherent acceleration.  The change in the ion gyroradius is linear in the wave frequencies and independent of wave amplitude.  The period of coherent motion is inversely proportional to the square of the wave amplitudes and is proportional to the fourth power of the wave frequency $\om$ ($\om\sim\om_1\sim\om_2$).  Furthermore, the deviation from resonance $\De\om=\om_1-\om_2-N\om_{ci}$ for which appreciable coherent acceleration occurs scales like $\om^{-4}$, while the range in $\De k_z=k_{1z}-k_{2z}$ for coherent motion scales like $\om^{-3}$.  This implies that for lower-frequency waves coherent ion acceleration is faster and less sensitive to small changes in wave parameters.

Coherent ion energization occurs for two waves with appropriately chosen frequencies.  An experiment is being constructed that will be able to test the theoretical predictions of this paper \cite{choueiri_aps_00}.  Coherent acceleration could also occur for a broadband spectrum of waves extending over at least two $\om_{ci}$ in frequency.  Such a situation can occur naturally in the Earth's ionosphere \cite{abhayjgr}.  Detailed analyses of a broad spectrum of waves, and of the effects of weak collisions, remain to be carried out in future work.

\begin{acknowledgments}
The authors thank Prof. A. Brizard for helpful discussions on the Lie transformation technique.  We also appreciate enlightening comments from Dr. D. B\'enisti about his work on this problem.  We thank R. Spektor for discussing his work with us and exploring possible experimental realizations of this process.  This work was supported by DOE Contract \mbox{DE-FG02-91ER-54109}, DOE/NSF Contract \mbox{DE-FG02-99ER-54555}, NSF Contract \mbox{ATM-0114462}, and Princeton University Subcontract \mbox{150-6804-1}.  DJS was partly supported by an NDSEG Graduate Fellowship.
\end{acknowledgments}

\appendix
\section{Lie Perturbation Method for Two Oblique Waves}
\label{app:lie}

We develop the Lie perturbation method following Refs.~\cite{lichtlieb} and \cite{cary} and follow the notation in Section~2.5 of Ref.~\cite{lichtlieb}.

The Lie method provides a Hamiltonian $\Hbar$ that describes just the coherent motion, and a change of coordinates that accounts for the incoherent fluctuations.  The physical variables $x=(q,p)$ are governed by the full Hamiltonian $H(x)$, and the new coordinates $\xbar=(\qbar,\pbar)$ are governed by $\Hbar(\xbar)$.  $\xbar$ depends on $x$ and a parameter $\ep$ which orders the perturbation via
\begin{equation}
\label{eqn:lietrans}
\frac{\p\xbar}{\p\ep}=[\xbar,w(\xbar,t)]_\xbar, \qquad \xbar(\ep=0)=x
\end{equation}
where $[f,g]_x=\sum_i[(\p f/\p q_i)(\p g/\p p_i)-(\p f/\p p_i)(\p g/\p q_i)]$ is the Poisson bracket.  The old coordinates enter only as a condition for $\ep=0$, which ensures that the transformation for any $w$ is canonical and near-identity.

The operator $T$ relates the representation of a physical quantity $f$ in the two coordinate systems by $f(\xbar)=(Tf)(x)$.  In particular, $f(\xbar)=\xbar$ gives $\xbar=Tx$.  $T$ satisfies
\begin{equation}\label{eqn:dTde}
\frac{\p T}{\p\ep}f(x)=-T[w(x,t),f(x)]_x
\end{equation}
$\Hbar$ is given by
\begin{equation}\label{eqn:Hbar}
\Hbar(\xbar) = T^{-1}H(x)+T^{-1}\int_0^\ep d\ep'T(\ep')\frac{\p w(x,t)}{\p t}
\end{equation}
The second term is not needed for an autonomous system.

We expand $w,H,T,$ and $\Hbar$ in powers of $\ep$ and equate terms at each order in $\ep$.  Collecting terms in (\ref{eqn:Hbar}) at each order in $\ep$ gives equations for $w_i$.  Upon carrying out the perturbation expansion to second order in $\ep$, we find
\begin{eqnarray} 
D_0w_1 &=& \Hbar_1-H_1    \label{eqn:D0eqn1}
\\ D_0w_2 &=& 2(\Hbar_2-H_2)-[w_1,\Hbar_1+H_1]    \label{eqn:D0eqn2}
\end{eqnarray}
$D_0f\equiv\p_tf+[f,H_0]$ is the time derivative along the unperturbed trajectories.  All expressions here are functions of the same set of coordinates.  For simplicity we use $x$ for this purpose, but the final expression for $\Hbar$ governs the evolution of $\xbar$.  Clearly, $\Hbar_0=H_0$.

For $T$ to be a near-identity operator, $w$ must remain small.  We choose $\Hbar_i$ in the right-hand side of (\ref{eqn:D0eqn1}) and (\ref{eqn:D0eqn2}) to eliminate any terms that would violate this condition.  Such terms are referred to as ``resonant'' terms.

For the two-wave problem, $H_0$ and $H_1$ are given in (\ref{eqn:H0H1}), while $H_i=0$ for $i\geq2$.  Using a Bessel-function identity (see p. 361 of \cite{absteg}), we obtain
\begin{equation}
H_1 = \sum_i\sum_{m=-\infty}^\infty \ep_i J_{m,i}\cos\psi_{mi}
\end{equation}
where $J_{m,i}\equiv J_m(k_{ix}\rho)$ and $\psi_{mi}\equiv m\phi+k_{iz}z-\nu_it+\al_i$.  Then from (\ref{eqn:D0eqn1})
\begin{equation} \label{eqn:dw1}
(\p_t+\p_\phi+v_z\p_z)w_1= \bar{H}_1-\sum_{i,m}\ep_iJ_{m,i}\cos\psi_{mi}
\end{equation}

The unperturbed orbits are $\phi=t+\phi_0,z=z_0,v_z=0,I=I_0$ (in a frame where the ion's initial $v_{z0}=0$).  Along these orbits there are no resonant terms on the right-hand side of (\ref{eqn:dw1}), so we choose $\Hbar_1=0$.  Then
\begin{equation} \label{eqn:w1}
w_1 = -\sum_{i,m}\frac{\ep_iJ_{m,i}}{m+k_{iz}v_z-\nu_i}\sin\psi_{mi}
\end{equation}
Since $H_2$ and $\Hbar_1$ are zero, (\ref{eqn:D0eqn2}) leads to
\begin{equation}
(\p_t+\p_\phi+v_z\p_z)w_2 = 2\Hbar_2 - [w_1,H_1]
\end{equation}
From (\ref{eqn:w1})
\begin{equation}\begin{split}
[w_1,H_1]=\sum_{i,j,m,n} \Big\{ &\frac{\ep_i\ep_j}{2\rho}\frac{1}{m-\mu_i}(-k_{jx}mJ_{m,i}J_{n,j}'+k_{ix}nJ_{m,i}'J_{n,j})\cos\Ga_+
\\&+\frac{\ep_i\ep_j}{2\rho}\frac{1}{m-\mu_i}(-k_{jx}mJ_{m,i}J_{n,j}'-k_{ix}nJ_{m,i}'J_{n,j})\cos\Ga_-
\\&+\thalf\ep_i\ep_jk_{iz}k_{jz}\frac{J_{m,i}J_{n,j}}{(m-\mu_i)^2}(-\cos\Ga_+ +\cos\Ga_-) \Big\}
\end{split}\end{equation}
where $\Ga_\pm\equiv \psi_{mi} \pm \psi_{nj}$.  Along the unperturbed orbits,  $\Ga_\pm=\{m-\nu_i\pm(n-\nu_j)\}t$ + const.  Some terms are resonant when $i=j$ regardless of the $\nu_i$'s.  Other terms are resonant when either $2\nu_i, N_+\equiv(\nu_1+\nu_2)$, or $N\equiv(\nu_1-\nu_2)$ is an integer.  We construct $\Hbar_2$ to cancel these terms:
\begin{equation}\begin{split}
\Hbar_2 = S_0(I,v_z) &+ \de_-S_-(I,v_z)\cos((\nu_1-\nu_2)(\phi-t)+(k_{1z}-k_{2z})z+\al_1-\al_2)
\\          &+ \de_+S_+(I,v_z)\cos((\nu_1+\nu_2)(\phi-t)+(k_{1z}+k_{2z})z+\al_1+\al_2)
\\          &+ \sum_i\de_iS_i(I,v_z)\cos(2\nu_i(\phi-t)+2k_{iz}z+2\al_i)
\end{split}\end{equation}
where $\de_-,\de_+,$ and $\de_i$ are unity when, respectively, $N,N_+,$ and $2\nu_i$ are integers and 0 otherwise.  Equations (\ref{eqn:S0}) and (\ref{eqn:Sm}) give $S_0$ and $S_-$, and
\begin{align}
S_+ =& -\sum_m \Big\{ \frac{\ep_1\ep_2}{4\rho(m-\mu_1)} (k_{1x}(m-N_+)J_{m,1}'J_{-m+N_+,2}+k_{2x}mJ_{m,1}J_{-m+N_+,2}') \notag
\\ & + \tfourth k_{1z}k_{2z}\ep_1\ep_2 \frac{J_{m,1}J_{-m+N_+,2}}{(m-\mu_1)^2} \notag
\\ & + \textrm{the same with subscripts 1 and 2 switched}  \Big\}
\\S_i =& -\sum_m \Big\{ \frac{\ep_i^2}{4\rho(m-\mu_i)} (mk_{ix}J_{m,i}J_{-m+2\nu_i,i}'+(m-2\nu_i)k_{ix}J_{m,i}'J_{-m+2\nu_i,i}) \notag
\\   & + \tfourth\ep_i^2k_{iz}^2\frac{J_{m,i}J_{-m+2\nu_i,i}}{(m-\mu_i)^2} \Big\}
\end{align}

The coherent Hamiltonian is
\begin{equation}
\Hbar(\xbar,t) = H_0(\xbar) + \Hbar_2(\xbar,t)
\end{equation}
Using $\psibar=\phibar-t$ as the coordinate conjugate to $\Ibar$, the transformed Hamiltonian is
\begin{equation} \label{eqn:htilapp}
\Htil(\psibar,\zbar,\Ibar,\vbar_z) = \thalf \vbar_z^2 + \Hbar_2
\end{equation}
$\Htil$ is a constant of the motion.

When only one resonance condition is satisfied, we find a second constant of the motion besides $\Htil$, which relates $\Ibar$ and $\vbar_z$ (when only $N$ is an integer, this constant is $\vbar_z-(\De k_z/N)\Ibar$). The dynamical system described by (\ref{eqn:htilapp}) is thus completely integrable.  When any two resonance conditions are satisfied it is easy to see that $\nu_1$ and $\nu_2$ must both be half-integers.  Then all four resonance conditions are satisfied.  It does not appear that, in this case, there exists a second constant of the motion.  The dynamics described by $\Htil$ could be stochastic.

To find the transformation relating $x$ and $\xbar$, we expand (\ref{eqn:dTde}) and use $\xbar=Tx$.  To first order in $\ep$ we obtain
\begin{equation}
\xbar = Tx \approx x - \ep[w_1(x,t),x]_x + O(\ep^2)
\end{equation}
As desired, the coordinate change is near-identity.  The relation between $I$ and $\Ibar$ is given in (\ref{eqn:ItoIbar}).

\section{Asymptotic forms for $S_0$ and $S_-$}
\label{app:asym}
Here we derive the asymptotic forms for the terms in $\Htil$, given in (\ref{eqn:Htil}), using results in Refs.~\cite{absteg,gradryz,chia1,spektor_iepc01}.  For $k_{1x}=k_{2x}=1$, let
\begin{eqnarray}
S_{0x}   &=& \sum_{i=1}^2 \ep_i^2 s_x(\rho,\mu_i,0)
\\S_{-x} &=& \ep_1\ep_2(s_x(\rho,\mu_1,N)+s_x(\rho,\mu_2,-N))
\\S_{0z} &=& \sum_{i=1}^2 k_{iz}^2\ep_i^2 s_z(\rho,\mu_i,0)
\\S_{-z} &=& \ep_1\ep_2k_{1z}k_{2z}(s_z(\rho,\mu_1,N)+s_z(\rho,\mu_2,-N))
\end{eqnarray}
where
\begin{eqnarray}
s_x(\rho,\mu,n)   &=& (-)^n\frac{\pi}{8}\csc\mu\pi(J_{\mu+1}J_{-(\mu+1)+n}-J_{\mu-1}J_{-(\mu-1)+n})
\\s_z(\rho,\mu,n) &=& (-)^{n+1}\frac{\pi}{4}\frac{\p}{\p\mu}\lb \csc\mu\pi J_\mu J_{-\mu+n} \rb
\end{eqnarray}
and $J_\mu=J_\mu(\rho)$.

Bessel functions of negative order are replaced with
\begin{equation}
J_{-\mu}\approx-\sin(\mu\pi)Y_\mu
\end{equation}
where $Y_\mu$ is the Bessel function of the second kind.  Using the asymptotic forms for $J_\mu$ and $Y_\mu$ \cite{absteg}, we obtain
\begin{eqnarray}
\si(\mu,\rho,n) &\equiv& J_\mu(\rho) Y_{\mu-n}(\rho) \sim \beta e^\ga
\\ \beta   &=& -\frac{1}{\pi}(\mu(\mu-n)\tanh\al_0\tanh\al_n)^{-1/2}
\\ \ga &=& \mu(\tanh\al_0-\al_0)-(\mu-n)(\tanh\al_n-\al_n)
\\ \sech\al_n &=& \frac{\rho}{\mu-n}
\end{eqnarray}

For $N+1\ll\mu$ all the $n/\mu$'s are small, and to leading order in $n/\mu$ we find
\begin{eqnarray}
\beta  &\approx& -\frac{1}{\mu\pi}(1-\xi^2)^{-1/2}
\\ \ga &\approx& -n\al_0
\\ \si &\approx& -\frac{1}{\mu\pi}\si_0(\xi,n)
\\ \si_0(\xi,n) &\equiv& (1-\xi^2)^{-1/2}\lp\frac{\xi}{1+\sqrt{1-\xi^2}}\rp^n
\end{eqnarray}

Thus,
\begin{eqnarray}
s_z &\approx& \frac{\pi}{4}\frac{\p}{\p\mu}\si(\mu,\rho,n)
\\ &\approx& \frac{f(\xi,n)}{\mu^2}
\\ f(\xi,n) &\equiv& \frac{1}{4}\lp\si_0+\xi\frac{\p\si_0}{\p\xi}\rp
\end{eqnarray}
Similarly,
\begin{eqnarray}
s_x &\approx& \frac{\pi}{8}(\si(\mu+1,\rho,n)-\si(\mu-1,\rho,n))
\\ &\approx& \frac{\pi}{8}(\si_1(\ep)-\si_1(-\ep))
\\ \si_1(\ep) &=& \frac{\si_0(\xi/(1+\ep),n)}{\mu(1+\ep)}
\end{eqnarray}
where $\ep\equiv1/\mu$ is small for $\mu\gg1$.  Expanding $\si_1$ to leading order in $\ep$,
\begin{eqnarray}
s_x &\approx& \frac{\pi}{4}\ep\si_1'(0)
\\ &=& \frac{f(\xi,n)}{\mu^2}
\end{eqnarray}
Thus,
\begin{equation}
s_x  \approx s_z \sim \frac{1}{\mu^2}
\end{equation}
$S_0$ and $S_-$ both scale like $1/\nu_1^2$ when $\nu_1\approx\nu_2$.

\newpage
\bibliography{035307PHP}

\newpage

\end{document}